\def\year{2021}\relax
\documentclass[letterpaper]{article} 
\usepackage{aaai21}  
\usepackage{times}  
\usepackage{helvet} 
\usepackage{courier}  
\usepackage[hyphens]{url}  
\usepackage{graphicx} 
\usepackage{natbib}  
\usepackage{caption} 
\usepackage{graphicx}
\usepackage{algorithm}
\usepackage{algorithmic}
\usepackage{amsmath,amssymb,amsfonts}

\usepackage{multirow}
\usepackage{graphbox}
\usepackage{hyperref}
\usepackage{textcomp}
\usepackage{xcolor}
\usepackage{flushend}
\usepackage{booktabs}

\title{Deep Kernel Supervised Hashing for Node Classification in Structural Networks }
\author{
Jia-Nan Guo, Xian-Ling Mao, Shu-Yang Lin, Wei Wei and Heyan Huang\\
Beijing Institute of Technology, Beijing, China\\
Huazhong University of Science and Technology, China\\
\{guojn, maoxl, hhy63\}@bit.edu.cn\\
linshuyang2017@gmail.com\\
Weiw@hust.edu.cn\\
}

\usepackage[switch]{lineno}  

\begin{document}

\maketitle
\begin{abstract}

Node classification in structural networks has been proven to be useful in many real world applications. With the development of network embedding, the performance of node classification has been greatly improved. However, nearly all the existing network embedding based methods are hard to capture the actual category features of a node because of the linearly inseparable problem in low-dimensional space; meanwhile they cannot incorporate simultaneously network structure information and node label information into network embedding. To address the above problems, in this paper, we propose a novel Deep Kernel Supervised Hashing (DKSH) method to learn the hashing representations of nodes for node classification. Specifically, a deep multiple kernel learning is first proposed to map nodes into suitable Hilbert space to deal with linearly inseparable problem. Then, instead of only considering structural similarity between two nodes, a novel similarity matrix is designed to merge both network structure information and node label information. Supervised by the similarity matrix, the learned hashing representations of nodes simultaneously preserve the two kinds of information well from the learned Hilbert space. Extensive experiments show that the proposed method significantly outperforms the state-of-the-art baselines over three real world benchmark datasets.

\end{abstract}

\section{Introduction}

Networks are ubiquitous in the real world, and many real-world datasets take the form of networks such as social networks, citation networks, language networks and biological networks. Generally, networks can be divided into two categories: structural networks \cite{wang2016structural,xie2019tpne} and attributed networks \cite{chen2020relation,yu2020structured}. Compared with attributed networks, structural networks are more widely used, which can be constructed by the most fundamental network structure information without using auxiliary information from node attributes. 

In structural networks, node classification is one of the most typical learning tasks, which focuses on exploiting the node interactions to predict the missing labels of unlabeled nodes in a structural network. Many real world applications can be modeled as the node classification problem, such as profession identification \cite{tu2017prism} and persona classification \cite{kaul2020persona}.

Generally speaking, existing node classification methods can be divided into two categories: traditional methods \cite{neville2000iterative,yamaguchi2015omni} and network embedding based methods \cite{tang2015line,grover2016node2vec,xie2019sim2vec,dai2019adversarial, xie2019tpne}. Compared with the traditional methods that directly infer posterior distribution of node labels from neighborhood information, network embedding based methods can achieve better performance by alleviating the curse of dimensionality for large-scale structural networks and avoiding cascading errors. However, nearly all the existing embedding based node classification methods in structural networks suffer from the following two problems: (1) They are hard to capture the actual category features hidden in highly nonlinear network structure, because of the linearly inseparable problem in low-dimensional space; (2) They only preserve network structure information into network embedding, without node label information.








To address the above problems, we propose a novel Deep Kernel Supervised Hashing, called DKSH, to learn node representations for node classification in structural networks. Specifically, a deep multiple kernel learning is first proposed to map nodes into suitable Hilbert space, which can deal with linearly inseparable problem of category features. Then, instead of only considering structural similarity and  ignoring category similarity between two nodes, a novel similarity matrix is designed to merge both network structure information and node label information. Supervised by the similarity matrix, the learned hashing representations of nodes can simultaneously preserve the two kinds of information from the learned Hilbert space. Extensive experiments show that the proposed method significantly outperforms the state-of-the-art baselines over three real world benchmark datasets.

The main contributions of our work are summarized as follows:

\begin{itemize}

\item We design a deep kernel hashing to maps nodes into suitable Hilbert space, which can deal with linearly inseparable problem of category features so as to generate good-quality hashing representations of nodes for node classification.

\item We define a novel similarity matrix in network embedding area to merge both network structure information and node label information. Supervised by the similarity matrix, the proposed method can incorporate simultaneously the two kinds of information into network embedding.

\item Extensive experiments over three real world benchmark datasets show that the proposed method significantly outperforms the state-of-the-art baselines.

\end{itemize} 

The rest of paper is arranged as follows: In Section 2, we will first review the related work of node classification and kernel hashing. Then, the details of our DKSH will be presented in Section 3. Moreover, in Section 4, we will present the experimental results of node classification over three real world benchmark datasets. Finally, the conclusions will be given to summary our work in Section 5.

\section{Related Work}

\subsection{Node Classification in Structural Networks} 

In this section, we discuss the recent trends and some state-of-the-art node classification methods in structural networks instead of attributed networks. Generally, existing node classification methods can be divided into two categories: traditional methods and embedding based methods. 

Traditional methods \cite{neville2000iterative,yamaguchi2015omni} pose node classification as an inference in an undirected Markov network, and then use iterative approximate inference algorithms to directly compute the posterior distribution of labels given the network structure. For example, OMNI-Prop \cite{yamaguchi2015omni} assigns each node with the prior belief about its label and then updates the label using the evidence from its neighbors, i.e., if the most of neighbors have the same label, then the rest also have the same label. However, these methods have a high computational complexity, which suffer from the curse of dimensionality for large-scale structural networks; meanwhile they cannot avoid cascading errors.




Different from traditional methods, network embedding based methods learn a classifier from the learned low-dimensional node representations, which can achieve better performance by alleviating the curse of dimensionality for large-scale structural networks and avoiding cascading errors. Nowadays, network embedding based methods become the recent trend for node classification. Essentially, this type of methods adopt following three steps \cite{cui2018survey}: (1) a network embedding algorithm, such as deep neural network \cite{perozzi2014deepwalk,wang2016structural} and matrix factorization \cite{ou2016asymmetric}, is applied to learn low-dimensional node representations with preserving rich network structure information; (2) the nodes with known labels are used as the training set; (3) a classifier, such as support vector classifier \cite{dai2019adversarial} and logistic regression classifier \cite{grover2016node2vec}, is learned from the representations and labels of training nodes to perform node classification. The representative methods include DeepWalk \cite{perozzi2014deepwalk}, node2vec \cite{grover2016node2vec}, SDNE \cite{wang2016structural} and Dwns \cite{dai2019adversarial}. DeepWalk first adopts random walk to extract local structure information of a node into node representation and then use an one-vs-rest logistic regression for classification. node2vec adopts a flexible method for sampling node sequences to strike a balance between local and global structure information in network embedding process and then also use an one-vs-rest logistic regression classifier to classify. SDNE first adopts a deep autoencoder to simultaneously extract both the first-order and the second-order similarity into node representations and then uses a support vector classifier for classification. Dwns improves DeepWalk with generative adversarial networks (GANs) based regularization methods to generate better node representation and then also use a support vector classifier to classify. With these methods, the performance of node classification has been greatly improved.

The major problems of previous methods are that: (1) They are hard to capture the actual category features of a node because of the linearly inseparable problem; (2) They only preserve network structure information into network embedding without considering node labels. Among these methods, node2hash \cite{wang2018feature} is the closest to our DKSH, which also uses a kernel hashing method to obtain node representations from network information. However, their algorithm adopts a shallow kernel hashing method \cite{shi2009hash}, which is still suffer from the above two problems. In contrast, the proposed method adopts a deep kernel supervised hashing, which can address the above two problems well. Before introducing our deep kernel supervised hashing, the existing kernel hashing methods is briefly described in the next subsection.

\begin{figure*}[t]
\centering
\includegraphics[width=15.5cm,height=8.2cm]{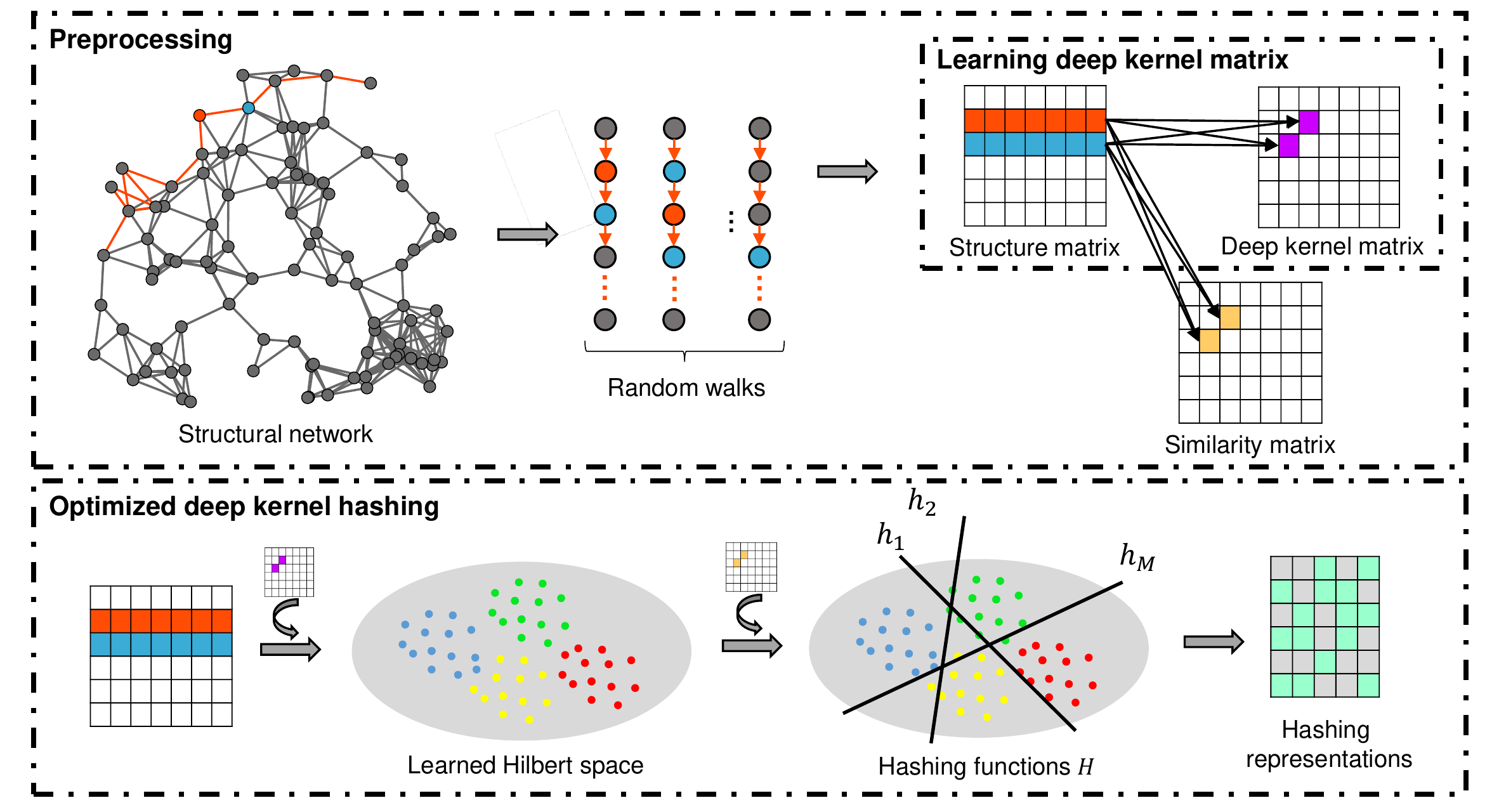}
\caption{The conceptual framework of Deep Kernel Supervised Hashing (DKSH). This framework contains three parts: (1) Preprocessing: extracting structure matrix from bold red random walk and then deriving similarity matrix; (2) Learning deep kernel matrix: learning DKL constructed by the structure matrix; (3) Optimized deep kernel hashing: mapping nodes into learned Hilbert space and then learning hashing functions $H$ supervised by the similarity matrix.}
\label{fig1}
\end{figure*}

\subsection{Kernel Hashing}
Kernel hashing is an useful method for nonlinear data, which maps original data into suitable Hilbert space and then learns hashing representations from this space. Generally, the existing kernel hashing methods can be categorized into single kernel hashing \cite{he2010scalable,wang2018feature} and multiple kernel hashing \cite{liu2014multiple}. Compared with multiple kernel hashing that is designed for multiple features data, single kernel hashing is the most fundamental method in kernel hashing. Thus, we take single kernel hashing as an example to introduce kernel hashing.

Single kernel hashing is an useful method to deal with classification tasks, which can learn hashing functions to map data from Hilbert space to hashing space. The formulation of hashing functions is:  
\begin{equation}
    \begin{split}
        B_{mi} = h_{m}(X_i) = sign(V_m^T \varphi(X_i)-b_m)
    \end{split}
    \label{eq1}
\end{equation}
with
$$
V_m = \sum_{r=1}^R W_{rm} \varphi(X_r)
$$
where $h_{m}$ is m-th hashing code of data $X_i$, $\varphi$ is the function that maps original data to Hilbert space, and $b_m$ is the threshold scalar. Besides, $V_m$ is the m-th hyperplane vector in the Hilbert space, which is a linear weighted combination of $R$ landmarks, i.e., $X_r, r = 1, \cdots, R$, with weight matrix $W_{R \times M}$. Note that, the landmarks are cluster centers produced by clustering or random choosing.

According to the definition of kernel matrix $K_{ij}=\varphi(X_i)^T\varphi(X_j)$, Equation~\eqref{eq1} can be rewritten in a kernel form:
\begin{equation}
    \begin{split}
        B_{i} = sign(W^T K_i-b)
    \end{split}
    \label{eq2}
\end{equation}
where $W_{R \times M}$ is the weight matrix of $R$ landmarks, $K_i$ is i-th column of a designed kernel matrix $K_{R \times N}$. In this way, the formulation of hashing functions is obtained, which can be used to learn nonlinear features from many data modalities, especially image \cite{he2010scalable}.

Nevertheless, both single kernel hashing and multiple kernel hashing adopt shallow kernel \cite{zhuang2011two}, which is often powerless to capture the actual features from highly nonlinear network data.

\section{Deep Kernel Supervised Hashing}

In this section, we first describe the problem formulation of node classification in structural networks, and then introduce the details of the proposed Deep Kernel Supervised Hashing (DKSH) method. The conceptual framework of DKSH is shown in Figure~\ref{fig1}. 

\subsection{Problem Formulation}

Formally, an undirected network is denoted as $G=(V,E,Y)$, where $V=\{v_i\}_{i=1}^N$ represents the set of $N$ nodes, $E=\{e_{ij}\}_{i,j=1}^N$ represents the set of edges between two nodes and $Y$ denotes the labels set. For $v_i$ and $v_j$ are linked by an edge, $e_{ij} = 1$. Otherwise, $e_{ij} = 0$. Network hashing embedding is to learn a set of hash functions $H=\{h_m\}_{m=1}^M$, which are used to map each node in $G$ into a low-dimension hashing representation $B_i \in \{-1, 1\}^M$, $M$ is the dimension of hashing representations. 

Given the labeled node set $V_L$ and the unlabeled node set $V_N$, where each node $v_i \in V_L$ is associated with a label $y_i \in Y$ but not in another, our goal is to predict the missing labels of unlabeled nodes $V_N$ with the learned hashing representations $B$.

\begin{algorithm}[t] 
    \caption{Building Structure Matrix} 
    \label{alg:Framwork} 
    \begin{algorithmic}[1] 
    \REQUIRE Network denoted as $G=(V, E, Y)$, window size $p$, walk length $l$ and walks per nodes $\gamma$.
    \ENSURE Structure matrix $P$.
    \STATE Initialize structure matrix $P=O$, $paths = \emptyset$.
    \STATE {\bf for} t = 0 to $\gamma$ {\bf do}
    \STATE ~~~~Nodes = Shuffle($V$)
    \STATE ~~~~{\bf for} each $v_i$ $\in$ Nodes {\bf do}
    \STATE ~~~~~~~~$path_{v_i}$ = RandomWalk(G, $v_i$, $l$)
    \STATE ~~~~~~~~Put $path_{v_i}$ in $paths$
    \STATE ~~~~{\bf end for}
    \STATE {\bf end for}
    \STATE Store $paths$
    \STATE {\bf for} each $path$ in $paths$ {\bf do}
    \STATE ~~~~{\bf for} each $v_i$ $\in$ $path$ {\bf do}
    \STATE ~~~~~~~~{\bf for} each $v_j$ $\in$ $path[i-p : i+p]$ {\bf do}
    \STATE ~~~~~~~~~~~~$P_{ij} = P_{ij}+(p+1-dis(v_i, v_j))/p$
    \STATE ~~~~~~~~{\bf end for}
    \STATE ~~~~{\bf end for}
    \STATE {\bf end for}
    \STATE Store structure matrix $P$
    \end{algorithmic}
    \label{alg1}
\end{algorithm} 

\subsection{Preprocessing Algorithms}

The algorithms of preprocessing is used to construct structure matrix by sampling network structure information and then similarity matrix by merging node labels and the structure matrix.

\subsubsection{Structure Matrix}
In network embedding area, the random walk is one of the most popular and powerful network sampling methods, which reflects the rich network structure information of center node in $G$. Generally, the relationship extracted from random walks contains $0$ and $1$, where $0$ is the relationship between unknown node pairs and $1$ is the relationship between similar node pairs \cite{perozzi2014deepwalk}. However, this type of relationship ignores the relative distance between center node and context nodes of it in window. Therefore, in this paper, we assign different weights to context nodes in window, according to their relative distance to the center node.

Initializing structure matrix $P=O$, where $O$ is zero matrix. For each similar node pairs $(v_i, v_j)$ in $window_{vi}^p$, where $v_i$ is the center node, $v_j$ is the context nodes of $v_i$ and $p$ is the window size, the recursive definition of $P$ is:
\begin{equation}
    \begin{split}
        P_{ij}^{\prime} = P_{ij} + \frac{p+1-dis(v_i, v_j)}{p}
    \end{split}
    \label{eq4}
\end{equation}
where $dis(v_i, v_j)$ is to compute the relative distance between $v_i$ and $v_j$ in the window. Note that, $(p+1-dis(v_i, v_j))/p$ is the weight provided by $(v_i, v_j)$, which is negatively related to the relative distance. In this way, $P_{ij}$ can reflect simultaneously the relative distance and the co-occurrence frequency of $(v_i, v_j)$ in random walks. More details of constructing structure matrix is shown in Algorithm~\ref{alg1}.

\subsubsection{Similarity Matrix}
According to DeepWalk \cite{perozzi2014deepwalk}, structure matrix reflects rich structure information, which can be treated as feature matrix. Thus, in order to simultaneously preserve network structure information and node label information, similarity matrix $S$ is defined as:
\begin{equation}
    \begin{split}
    S_{ij}=
    \left\{
    \begin{array}{lcl}
    \exp{(-\frac{\|P_i-P_j\|^2}{\max(dis^2)})}  &,  & {y_j = y_i}\\
    0     &,    & otherwise\\
    \end{array} 
    \right.
    \end{split}
    \label{eq5}
\end{equation}
where, $P_i$ and $P_j$ are feature vectors of $v_i$ and $v_j$, $\max(dis^2)$ is the max globally distance between all the feature vectors and $y_i$ is the label of node $v_i$.

\subsection{Learning Deep Kernel Matrix}
Learning a deep kernel matrix aims to map nodes into suitable Hilbert space, which detail the architecture and implementation of deep multiple kernel learning.

\subsubsection{Deep Multiple Kernel Learning}

Multiple kernel learning (MKL) \cite{liu2014multiple} is a widely used technique for kernel designing. Its principle consists in learning, for a given support vector classifier, the most suitable convex linear combination of standard elementary kernels. However, this kind of linear combination of kernels is a shallow way, which often cannot capture highly nonlinear features. In this way, deep multiple kernel learning (DKL) is proposed \cite{strobl2013deep,jiu2017nonlinear}. Interestingly, network structure information is just a highly nonlinear network information. Thus, in this section, we describe how to make DKL to fit network data.

\begin{figure}[t]
\centering
\includegraphics[width=7.75cm,height=4cm]{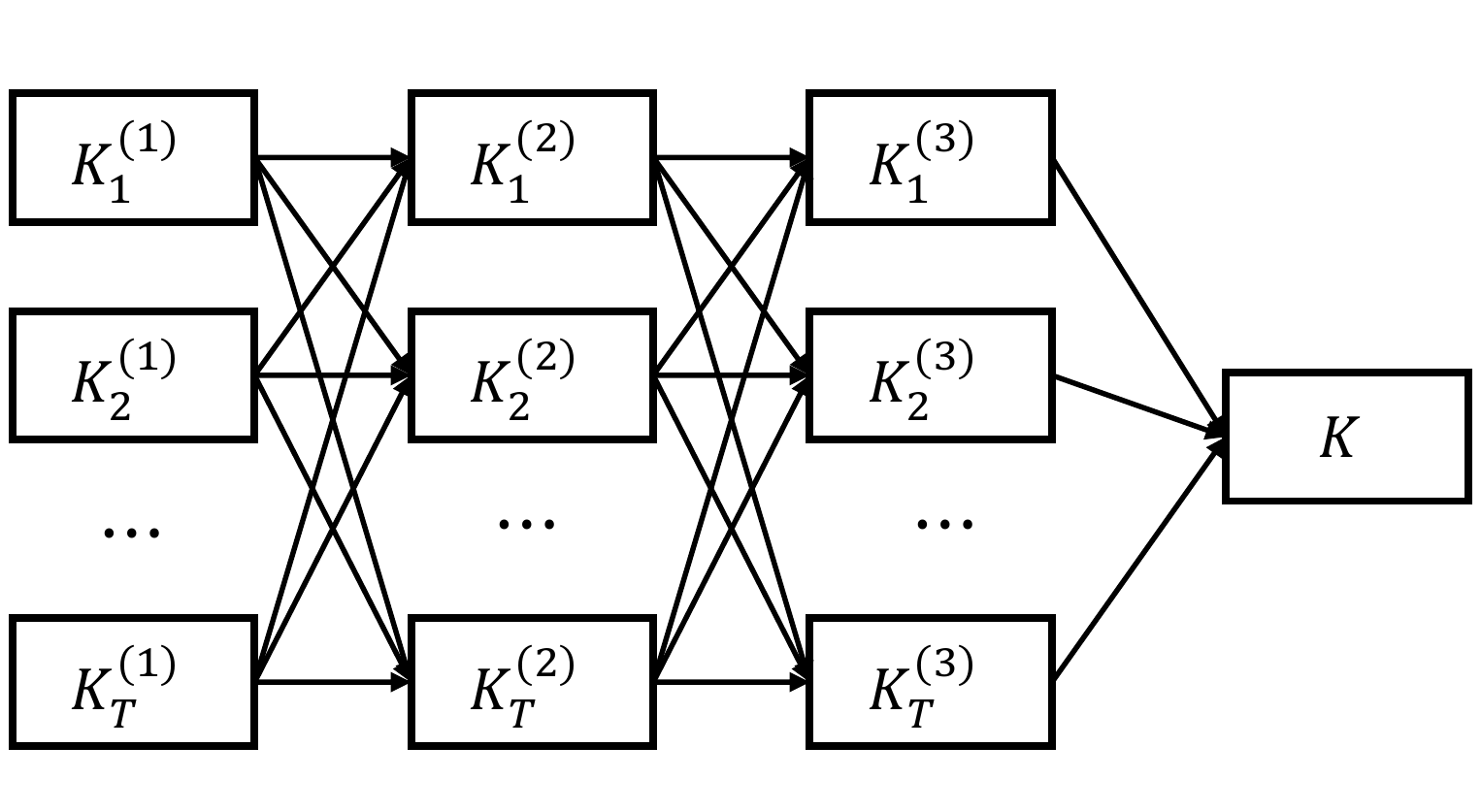}
\caption{The architecture of a 3-layers deep multiple kernel learning. Lines represent the weight for each kernel matrix, $\mu_{t}^{(l)}$.}
\label{fig2}
\end{figure}

Figure~\ref{fig2} shows the architecture of our DKL, which adopts a nonlinear multi-layered combination of multi-kernels. The recursive definition of our deep kernel is:
\begin{equation}
\begin{split}
        & K_{t}^{(l)} = g_t(\sum_{t=1}^T \mu_{t}^{(l-1)} K_t^{(l-1)})\\
        & \begin{array}{l@{\quad}l}
        s.t. & \mu \geq 0
        \end{array}  
\end{split}
\label{eq3}
\end{equation}
where $g_t(\cdot)$ is the activation function for kernel matrix $K_t$ like $rbf()$, which can map feature matrix (or kernel matrix) to kernel matrix. Moreover, we assume that the architecture of DKL has $L$ layers and each layer contains $T$ single kernel matrices, $l \in \{1, \cdots, L\}$, $t \in \{1, \cdots, T\}$. In this case, $K_{t}^{(l)}$ expresses the kernel matrix of l-th layer and t-th unit in this model. Besides, $K_{t}^{(1)} = g_t(P)$, where P is the structure matrix of a structural network (see Algorithm~\ref{alg1}), and the finial output of the proposed DKL is $K = \sum_{t=1}^T \mu_{t}^{(L)} K_t^{(L)}$. 


\begin{algorithm}[t] 
    \caption{Learning Deep Kernel Matrix} 
    \label{alg:Framwork} 
    \begin{algorithmic}[1]
    \REQUIRE Initial network parameters $\mu=\frac{1}{T}$, structure matrix $P$, the number of nodes $N$.
    \ENSURE Optimal $\mu$, coefficients of SVM $\alpha$.
    \STATE {\bf repeat}
    \STATE ~~~~Fix $\mu$, compute output deep kernel $K$.
    \STATE ~~~~$\alpha$ is optimized by SVM solver.
    \STATE ~~~~Fix $\alpha$, compute the gradient $\nabla_{\mu}$ of $\mathcal{T}_{span}$ w.r.t $K$.
    \STATE ~~~~Update weight $\mu$, according to $\nabla_{\mu}$, and keep $\mu \geq 0$.
    \STATE {\bf until} Convergence
    \end{algorithmic}
    \label{alg2}
\end{algorithm} 

\subsubsection{Implementation} 

In order to optimize the network parameters $\mu$ of the proposed DKL, we use the backward information from an one-vs-rest SVM classifier. The SVM classifier for network data is designed as $sgn(\sum_{i=1}^N \alpha_i y_i K(v_i, v)+b)$, where $y_i$ is the label of node $v_i$. Generally, $\mu$ can be optimized by minimizing a objective function, which is a global hinge loss of the classifier like \cite{jiu2017nonlinear}. However, in order to reduce the risk of over-fitting, we use span bound method to learn deep kernel \cite{liu2011learning}. Under the assumption that the set of support vectors remains the same during the leave-one-out procedure, the span bound can be stated as:
\begin{equation}
    \begin{split}
        \mathcal{L}((v_1, y_1), \cdots, (v_N, y_N)) \leq \sum_{i=1}^N \varphi(\alpha_i D_i^2 - 1)
    \end{split}
    \label{eq6}
\end{equation}
where $\mathcal{L}$ is leave-one-out error, and $D_i$ is the distance between the point $\varphi(v_i)$ and the set $\Gamma_i = \{\sum_{j \neq i, \alpha_j>0} \lambda_j \varphi(v_j)|\sum_{j \neq i} \lambda_j=1\}$. Based on Equation~\eqref{eq6}, we can formulate the objective function of deep multiple kernel learning as minimize the upper bounds of leave-one-out error:
\begin{equation}
    \begin{split}
        \min_{\mu, \alpha} \mathcal{T}_{span} =  \sum_{i=1}^N \varphi(\alpha_i D_i^2 - 1)
    \end{split}
    \label{eq7}
\end{equation}
Here, the objective function is optimized w.r.t two parameters: $\mu$ and $\alpha$. Alternating optimization strategy is adopted, i.e., we fix $\mu$ to optimize $\alpha$, and then vice-versa. At each iteration, when $\mu$ is fixed, the deep kernel $K$ is also fixed, $\alpha$ can be auto-optimized using a SVM classifier \cite{fan2008liblinear}. When $\alpha$ fixed, $\mu$ can be directly optimized by computing the gradient of Equation~\eqref{eq7} \cite{strobl2013deep}. The iterative procedure continues until convergence or when a maximum number of iterations is reached (see Algorithm~\ref{alg2}).

\subsection{Optimized Deep Kernel Hashing}



The function of deep kernel hashing is used to learn the hashing representations of nodes from the learned Hilbert space. Note that, in the section of related work, we introduce the existing kernel hashing methods. However, both single kernel hashing and multiple kernel hashing are shallow kernel hashing methods, which are often powerless to learn a suitable Hilbert space for highly nonlinear network structure. In order to address this problem, we extend shallow kernel into the learned deep kernel. According to Equation~\eqref{eq2} and Equation~\eqref{eq5}, the proposed deep kernel hashing functions can be written as:
\begin{equation}
    \begin{split}
        B_{i} = sign(W^T K_i-b)
    \end{split}
    \label{eq8}
\end{equation}
with
$$
K_{R \times N} = (\sum_{t=1}^T \mu_{t}^{(L)} K_t^{(L)})_{R \times N}
$$
where $W_{R \times M}$ is weighted matrix of landmarks, $K_i$ is i-th column of the learned deep kernel matrix $K_{R \times N}$. 

With the similarity matrix and the hashing functions, the form of similarity-distance product minimization \cite{wang2017survey} is adopted to design the following objective function for network hashing representation:
\begin{equation}
\begin{split}
    & \min\limits_{W} \quad \frac{1}{2}\sum\limits_{i,j=1}^N S_{ij}\|B_i-B_j\|^2 + \lambda \sum_{m=1}^M \|V_m\|^2\\
    & \begin{array}{l@{\quad}l}
    s.t.& \sum\limits_{i=1}^N B_i = 0\\ &\frac{1}{N}\sum\limits_{i=1}^N B_iB_i^T = I\\
\end{array}
\end{split}
\label{eq9}
\end{equation}
where $S$ is the similarity matrix. $B_i$ is the hashing representation of $v_i$ obtained from Equation~\eqref{eq8}, which has the same constraints. $\sum_{m=1}^M \|V_m\|^2$ is utilized to a regularized term to control the smoothness of hyperplane vector $V_m$. The constraint $\sum_{i=1}^N B_i = 0$ is to make sure bit balance, i.e., $50\%$ to be $1$ and $50\%$ to be $-1$. The constraint $\frac{1}{N}\sum_{i=1}^N B_iB_i^T = I$ is to ensure bit uncorrelation. In this way, we can obtain compact hashing representations of nodes.

Using Laplacian matrix $L = diag(S\mathbf{1})-S$, the objective function can be derived as:
\begin{equation}
\begin{split}
    & \min\limits_{W} \quad tr(W^T \frac{(C+C^T)}{2} W)\\
    & \begin{array}{l@{\quad}l}
    s.t.& W^T G W=I
    \end{array}
\end{split}
\label{eq10}
\end{equation}
with
$$
C = K_{R \times N} L K_{R \times N}^T + \lambda K_{R \times R}
$$
$$
G = \frac{1}{N}K_{R \times N}(I-\frac{1}{N}\mathbf{1}\mathbf{1}^T)K_{R \times N}^T
$$
Here, $b = -(\frac{1}{N})W^TK_{R \times N} \mathbf{1}$, $W_{R \times M}$ is weighted matrix of landmarks. Note that, the derivation follows \cite{he2010scalable}.  

\subsubsection{Implementation}
For simpler implementation, Equation~\eqref{eq10} can be further rewritten into an eigen vector problem:
\begin{equation}
\begin{split}
    & \min\limits_{W} \quad tr(\hat W^T \hat C \hat W)\\
    & \begin{array}{l@{\quad}l}
    s.t.& \hat W^T \hat W=I
    \end{array}
\end{split}
\label{eq11}
\end{equation}
with
$$
\hat C = \Lambda^{-\frac{1}{2}} T^T \frac{(C+C^T)}{2} T \Lambda^{-\frac{1}{2}}
$$
$$
G = T_0 \Lambda_0 T_0^T
$$
$$
W = T \Lambda^{-\frac{1}{2}} \hat W
$$
where $\Lambda$ is a diagonal matrix consisting of $M$ largest elements of $\Lambda_0$, and $T$ is the corresponding columns of $T_0$. In this way, the solution of this eigen vector problem is matrix $\hat W$, which is $M$ eigen vectors of matrix $\hat C$. Given $\hat W$, $W$ can be directly obtained by $W = T \Lambda^{-\frac{1}{2}} \hat W$. Based on $W$, we can get the hashing representations of $v_i, i=1, \cdots, N$, according to Equation~\eqref{eq8}.

Note that, after obtaining the hashing representations of nodes, an off-the-shelf classifier is trained to predict the missing labels of unlabeled nodes.

\section{Experiments}
In the previous section, the proposed method incorporates simultaneously network structure information and node label information into the hashing representations of nodes. In this section, extensive node classification experiments are conducted to verify that the proposed method can improve the performance of node classification in structural networks. 

\begin{table*}[t!]
    \begin{center}
		\caption{ Accuracy ($\%$) of Node Classification over Wiki. }
		\label{tab2}
                \scalebox{1}{\begin{tabular}{l|ccccccccc}
				\hline \hline
				\multirow{2}{*}{Methods} & \multicolumn{9}{c}{\% Labeled Nodes} \\ \cline{2-10}
				& 10\%  & 20\% & 30\% & 40\% & 50\%  & 60\% & 70\% & 80\% & 90\% \\ \hline 

				DeepWalk   & 46.60 & 54.48 & 59.05 & 62.70 & 64.66 & 65.95 & 66.98 & \underline{68.37} & \underline{68.78} \\ 
				Line & 57.88 & 61.08 & 63.50 & 64.68 & \underline{66.29} & 66.91 & \underline{67.43} & 67.46 & 68.61 \\ 
				GraRep  & \underline{58.57} & \underline{61.91} & \underline{63.58} & 63.77 & 64.68 & 65.39 & 65.92 & 65.18 & 67.05 \\  
				nove2vec   & 55.94 & 59.67 & 61.11 & 64.21 & 65.08 & 65.58 & 66.76 & 67.19 & 68.73 \\  
				AIDW   & 57.32 & 61.84 & 63.54 & \underline{64.90} & 65.58 & 66.54 & 65.59 & 66.58 & 68.02 \\ 
				Dwns & 55.77 & 59.63 & 61.98 & 64.01 & 64.59 & \underline{66.99} & 66.45 & 67.55 & 67.51 \\ \hline
				node2hash & 53.35 & 55.32 & 57.74 & 59.65 & 61.28 & 60.96 & 62.83 & 62.08 & 64.07 \\ \hline
				\textbf{DKSH-1L} & 66.50 & 69.77 & 70.91 & 72.13 & 71.92 & 74.45 & 73.38 & 74.39 & 73.53 \\ 
				\textbf{DKSH-2L} & 66.65 & 70.72 & 72.70 & 72.70 & 74.06 & 75.07 & 74.13 & 74.05 & 75.85 \\ 
				\textbf{DKSH}  & \textbf{69.05} & \textbf{71.20} & \textbf{73.41} & \textbf{74.35} & \textbf{74.38} & \textbf{75.16} & \textbf{74.85} & \textbf{75.30} & \textbf{77.10} \\   \hline \hline 

		\end{tabular}}
    \end{center}
\end{table*}

\begin{table*}[t!]
    \begin{center}
		\caption{ Accuracy ($\%$) of Node Classification over Cora. }
		\label{tab3}
                \scalebox{1}{\begin{tabular}{l|ccccccccc}
				\hline \hline
				\multirow{2}{*}{Methods} & \multicolumn{9}{c}{\% Labeled Nodes} \\ \cline{2-10}
				& 10\%  & 20\% & 30\% & 40\% & 50\%  & 60\% & 70\% & 80\% & 90\% \\ \hline 

				DeepWalk   & 64.60 & 69.85 & 74.21 & 76.68 & 77.59 & 77.68 & 78.63 & 79.35 & 79.23 \\ 
				Line & 66.06 & 70.86 & 72.25 & 73.94 & 74.03 & 74.65 & 75.12 & 75.30 & 75.76 \\ 
				GraRep  & \underline{\textbf{74.98}} & 77.48 & 78.57 & 79.38 & 79.53 & 79.68 & 79.75 & 80.89 & 80.74 \\  
				nove2vec   & 73.96 & \underline{78.04} & \underline{80.07} & \underline{81.62} & \underline{\textbf{82.16}} & 82.25 & 82.85 & 84.02 & \underline{84.91} \\  
				AIDW   & 73.83 & 77.93 & 79.43 & 81.16 & 81.79 & \underline{\textbf{82.27}} & \underline{82.93} & \underline{84.11} & 83.69 \\ 
				Dwns & 73.20 & 76.98 & 79.83 & 80.56 & 82.27 & 82.52 & 82.92 & 82.97 & 84.54 \\ \hline
				node2hash & 55.99 & 56.16 & 63.87 & 67.51 & 70.24 & 70.20 & 71.34 & 72.83 & 73.43 \\ \hline
				\textbf{DKSH-1L} & 71.55 & 75.32 & 76.64 & 76.74 & 78.63 & 78.60 & 79.80 & 79.93 & 81.40 \\ 
				\textbf{DKSH-2L} & 72.90 & 77.83 & 79.40 & 79.57 & 80.99 & 81.14 & 81.16 & 80.22 & 82.73 \\ 
				\textbf{DKSH}  & 74.22 & \textbf{78.71} & \textbf{80.55} & \textbf{81.89} & 81.89 & 82.03 & \textbf{83.87} & \textbf{84.69} & \textbf{86.23} \\   \hline \hline 

		\end{tabular}}
    \end{center}
\end{table*}

\begin{table*}[t!]
    \begin{center}
		\caption{ Accuracy ($\%$) of Node Classification over Citeseer. }
		\label{tab4}
                \scalebox{1}{\begin{tabular}{l|ccccccccc}
				\hline \hline
				\multirow{2}{*}{Methods} & \multicolumn{9}{c}{\% Labeled Nodes} \\ \cline{2-10}
				& 10\%  & 20\% & 30\% & 40\% & 50\%  & 60\% & 70\% & 80\% & 90\% \\ \hline 

				DeepWalk   & 45.53 & 50.98 & 53.79 & 55.25 & 56.05 & 56.84 & 57.36 & 58.15 & 59.11 \\ 
				Line & 47.03 & 50.09 & 52.71 & 53.52 & 54.20 & 55.42 & 55.87 & 55.93 & 57.22 \\ 
				GraRep  & 50.60 & 53.56 & 54.63 & 55.44 & 55.20 & 55.07 & 56.04 & 55.48 & 56.39 \\  
				nove2vec  & \underline{50.78} & \underline{55.89} & \underline{57.93} & \underline{58.60} & 59.44 & 59.97 & 60.32 & 60.75 & 61.04 \\  
				AIDW  & 50.77 & 54.82 & 56.96 & 58.04 & \underline{59.65} & \underline{60.03} & \underline{60.99} & \underline{61.18} & \underline{62.84} \\ 
				Dwns & 50.00 & 53.74 & 57.37 & 58.59 & 59.00 & 59.53 & 59.62 & 59.51 & 60.18 \\ \hline
				node2hash & 38.58 & 47.96 & 49.29 & 50.96 & 52.66 & 52.30 & 53.22 & 56.11 & 57.23 \\ \hline
				\textbf{DKSH-1L} & \textbf{58.17} & 58.82 & 59.28 & 59.82 & 60.10 & 59.50 & 60.46 & 60.33 & 60.24 \\ 
				\textbf{DKSH-2L} & 57.44 & 58.80 & 59.72 & 59.01 & 59.73 & 59.67 & 60.26 & 61.84 & 61.75 \\ 
				\textbf{DKSH}  & 57.97 & \textbf{59.81} & \textbf{59.78} & \textbf{59.96} & \textbf{60.59} & \textbf{60.89} & \textbf{61.79} & \textbf{62.84} & \textbf{63.25} \\   \hline \hline
		\end{tabular}}
    \end{center}
\end{table*}

\subsection{Experimental Setup}
\subsubsection{Datasets}
We conduct experiments on three real world benchmark datasets, which are popularly used in many previous works \cite{dai2019adversarial,zhao2020deepemlan}. Wiki \cite{sen2008collective} is a network with nodes as web pages and edges as the hyperlinks between web pages. The network has 2,405 nodes, 17,981 edges, and 17 different labels. Cora \cite{mccallum2000automating} is a network of citation relationships of scientific papers. The network has 2,708 nodes, 5,429 edges, and 7 different labels. Citeseer is also a scientific paper citation network constructed by \cite{mccallum2000automating}. The network has 3,312 nodes, 4,732 edges, and 6 different labels. We regard these three networks as undirected structural networks, and do some preprocessing on the original datasets by deleting nodes with zero degree. 

\subsubsection{Metric}
Following previous works \cite{dai2017adversarial,dai2019adversarial}, we employ the popularly used \textbf{Accuracy} to evaluate the performance of node classification. In this paper, accuracy measures the micro-averaged accuracy of the aggregated contributions of all classes. 

\subsubsection{Baselines}
We compare our DKSH with the state-of-the-art baselines and its variants. For fair comparisons, all the selected baselines are widely used structural network embedding methods, which need not node attributes information. Besides, among these baselines, node2hash is the only hashing based method like our DKSH. The details of the baselines are as follows: 

\begin{itemize}
\item \textbf{DeepWalk} \cite{perozzi2014deepwalk}: DeepWalk is an unsupervised method, which uses local structure information obtained from truncated random walks to learn low-dimensional feature representations of nodes.
\item \textbf{Line} \cite{tang2015line}: Line uses the breadth-first strategy to sample the inputs, based on node neighbors. The method preserves both the first-order and second-order similarity in network embedding process.
\item \textbf{GraRep} \cite{cao2015grarep}: GraRep applies SVD technique to different k-step probability transition matrix to learn node representations, and finally obtains global representations through concatenating all k-step representations.
\item \textbf{node2vec} \cite{grover2016node2vec}: node2vec differs from DeepWalk by proposing more flexible method for sampling node sequences to strike a balance between local and global structure information.
\item \textbf{AIDW} \cite{dai2017adversarial}: AIDW is an inductive version of DeepWalk with GAN-based regularization methods. A prior distribution is imposed on node representations through adversarial learning to achieve a global smoothness in the distribution.
\item \textbf{Dwns} \cite{dai2019adversarial}: Dwns is also an inductive version of DeepWalk, which introduces a succinct and effective regularization technique, namely adversarial training method, in network embedding process.

\item \textbf{node2hash} \cite{wang2018feature}: node2hash uses the encoder-decoder framework, where the encoder is used to map the structural similarity of nodes into a feature space, and the decoder is used to generate the node representations through a single kernel hashing.
\end{itemize}
Besides, we also include two additional baselines, namely \textbf{DKSH-1L} and \textbf{DKSH-2L}, which are the variants of our DKSH and respectively represent DKSH with 1-layer MKL and 2-layers MKL. Note that, although graph neural networks based methods \cite{tang544triple,zhang2019graph} are relevant to our DKSH, to the best of our knowledge, nearly all these methods need node attributes information. It means that these methods are limited, and we do not select them as baselines. 

\subsubsection{Parameter Settings}
For DKSH and its variants including DKSH-1L and DKSH-2L, the window size $p$, walk length $l$, walks per nodes $\gamma$, number of kernel in each layer $T$, number of landmarks $R$ and regularization parameter $\lambda$ are respectively set to 50, 200, 10, 4, 256 and 0.0001. Different from DKSH-1L and DKSH-2L, DKSH adopts a 3-layer DKL. For each layers in these kernel models, we consider 4 elementary kernels, i.e., a linear kernel, an RBF kernel with $\gamma$ = 1, a sigmoid kernel with $\alpha = -1 \times 10^{-4}$ and $\beta = 1$, and a polynomial kernel with $\alpha = 1$, $\beta = 1$, and $\delta = 2$. Besides, the dimension of node representations are set to 128 for all methods, and the other parameters are set to be the default value for the baselines. 

\begin{figure*}[t!]
\centering
\includegraphics[width=12cm,height=4.2cm]{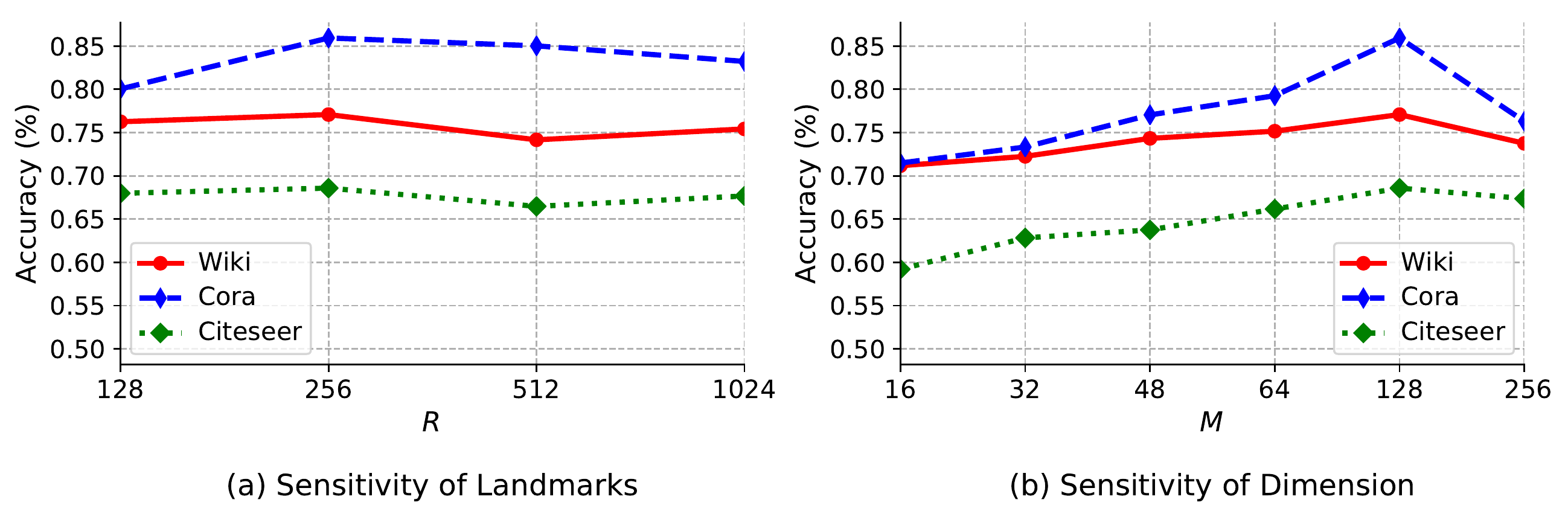}
\includegraphics[width=18cm,height=4.2cm]{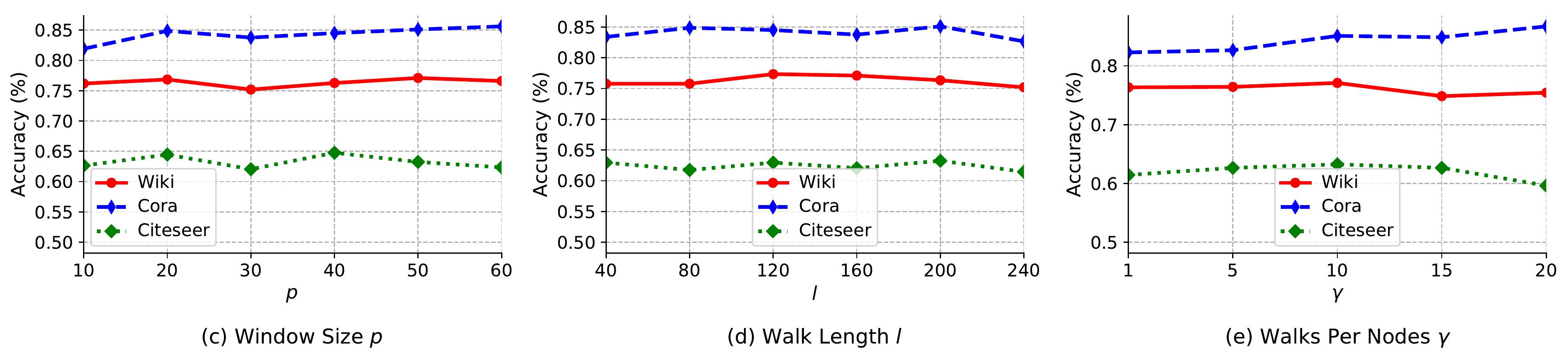}
\caption{Parameter sensitivity analysis of our DKSH over Wiki, Cora and Citeseer.}
\label{fig3}
\end{figure*}

\subsection{Comparison of Node Classification Performance}

For node classification, a portion of the labeled nodes is randomly selected as training data in which there is a same increment from 10\% to 90\% on training ratio and the rest of unlabeled nodes is used to test. Besides, for fair comparisons, all the methods adopt a support vector classifier in Liblinear \cite{fan2008liblinear}. Table~\ref{tab2} lists the results of node classification over Wiki. Table~\ref{tab3} lists the results of node classification over Cora. Table~\ref{tab4} lists the results of node classification over Citeseer. From these tables, it can be observed that:

\begin{itemize}

\item The proposed method DKSH consistently outperforms all the state-of-the-art baselines across all the ratio of labeled nodes over Wiki and Citeseer. As shown in Tables~\ref{tab2}, when the training ratio is 10\%, our DKSH performs better than these baselines with 90\% training ratio. As shown in Tables~\ref{tab3}, with the training ratio change from 10\% to 90\%, our DKSH outperforms these baselines about 1.5\%. Besides, over Cora, although the baselines can outperforms our DKSH with training ratio 10\%, 50\% and 60\%, our DKSH is still competitive.

\item DKSH consistently outperforms DKSH-2L, and DKSH-2L mostly outperforms DKSH-1L on three datasets across all the ratio of labeled nodes. For example, DKSH achieves about 1\% gain in Accuracy on Wiki, about 3\% gain in Accuracy on Cora, about 1.5\% gain in Accuracy on Citeseer with ratio of labeled nodes being 90\%. Besedes, DKSH-2L mostly gives about 1\% gain in Accuracy over DKSH-1L on Wiki, Cora and Citeseer. These results demonstrate that it is easier to capture the actual category features of a node in a more suitable Hilbert space.

\item It can significantly improve the performance of node classification by incorporating simultaneously network structure information and node label information into network embedding, instead of only network structure information. For example, both of node2hash and DKSH-1L adopt a single kernel hashing method. Compared with node2hash, DKSH-1L achieves more than 10\% gain in Accuracy on Wiki, more than 11\% gain in Accuracy on Cora, about 5\% gain in Accuracy on Citeseer across all training ratios. 

\item The hashing method does not necessarily lead to accuracy loss. In fact, it may avoid over-fitting in this paper. For example, both DKSH and node2hash use hashing methods to learn node representations. The results show that DKSH is better than nearly all the non-hashing methods, and node2hash has a competitive performance compared with DeepWalk and Line.

\end{itemize}

\subsection{Sensitivity Analysis}
In this subsection, we analyze sensitivity of parameters in our DKSH, i.e.,  the number of landmarks $R$, the dimension of node representations $M$, the window size $p$, walk length $l$ and walks per nodes $\gamma$ on Wiki, Cora and Citeseer datasets with ratio of labeled nodes being 90\% being 128bits. More specially, Figure \ref{fig3} (a) shows the affect of the parameter $R$ over the three datasets with the value between 128 and 1,024. Figure \ref{fig3} (b) shows the affect of the parameter $M$ over the three datasets with the value between 16 and 256. Figure \ref{fig3} (c) shows the affect of the parameter $p$ over the three datasets with the value between 10 and 60. Figure \ref{fig3} (d) shows the affect of the parameter $l$ over the three datasets with the value between 40 and 240. Figure \ref{fig3} (e) shows the affect of the parameter $\gamma$ over the three datasets with the value between 1 and 20. Note that except for the parameter being tested, all the other parameters are set to default values.


Figure~\ref{fig3} (a) shows Accuracy of our DKSH w.r.t. the number of landmarks $R$. When the dimension increases from $M$ to $1,024$, the Accuracy relatively stable over the three benchmark datasets, which means our DKSH is not sensitive on the number of landmarks $R$. However, training time is positively correlated with R. Thus, we select an relatively small $R$, i.e., 256.

Figure~\ref{fig3} (b) shows Accuracy of our DKSH w.r.t. the dimension of node representations $M$. DKSH achieves the best Accuracy over the three benchmark datasets when $M=128$. The reason why the Accuracy over the three benchmark datasets improves first when $M$ varies from 16 to 128 is largely because more features are captured into higher dimension of node representation. However, when the dimension further increases to 256, the linear SVM classifier is powerless to classify these high-dimensional node representations.

Figure~\ref{fig3} (c), (d) and (e) show Accuracy of our DKSH w.r.t. the window size $p$, walk length $l$ and walks per nodes $\gamma$. It can be found that DKSH is not sensitive to all the three parameters, i.e., $p$, $l$ and $\gamma$. For instance, DKSH can achieve good performance on all the three datasets in the range of 10 to 60 for the parameter $p$, and also can achieve good performance on all the three datasets with $40 \leq l \leq 240$. Furthermore, DCHUC can get the high MAP values with only one walks per nodes.

\section{Conclusions}
In this paper, we propose a novel Deep Kernel Supervised Hashing (DKSH) method to learn the hashing representations of nodes for node classification in structural networks. In DKSH, we designed a deep kernel hashing to mapping highly non-linear network structure information into suitable Hilbert space to deal with linearly inseparable problem, and define a novel similarity to incorporate simultaneously network structure information and node label information into network embedding. The experimental results demonstrate the superior usefulness of the proposed method in node classification.
 
\bibliography{DKSH}
\bibliographystyle{aaai21}
\end{document}